\documentclass[twocolumn,letter]{jpsj2} 
\title{Novel Charge Order and Superconductivity in Two-Dimensional Frustrated Lattice at Quarter Filling}

\author{Hiroshi \textsc{Watanabe}\thanks{E-mail address: hwatanabe@hosi.phys.s.u-tokyo.ac.jp} and Masao \textsc{Ogata}}

\inst{Department of Physics, University of Tokyo, Hongo, Bunkyo-ku, Tokyo 113-0033, Japan}

\abst{Motivated by the various physical properties observed in $\theta$-(BEDT-TTF)$_2$X, we study the ground state of extended Hubbard model
on two-dimensional anisotropic triangular lattice at 1/4-filling with variational Monte Carlo method. It is shown that the nearest-neighbor 
Coulomb interaction enhances the charge fluctuation and it induces the anomalous state such as charge-ordered metallic state and the triplet
next-nearest-neighbor $f$-wave superconductivity. We discuss the relation to the real materials and propose the unified view of the family 
of $\theta$-(BEDT-TTF)$_2$X.}

\kword{charge order, charge fluctuation, superconductivity, organic conductor, $\theta$-(BEDT-TTF)$_2$X, variational Monte Carlo method}

\begin{document} 
\maketitle
Charge order (CO) is one of the important phenomena in strongly correlated electron systems as well as superconductivity and magnetism.
In a charge-ordered system, the cooperation or competition between charge and spin degrees of freedom cause the various interesting 
properties as seen in manganese oxides,~\cite{Tomioka,Tokura} vanadium oxides,~\cite{Isobe} high-$T_{\mathrm{c}}$ cuprates~\cite{Tranquada}
and organic conductors.~\cite{HMori1,Takano} 
Above all, the family of organic conductors $\theta$-(BEDT-TTF)$_2$X (we denote $\theta$-(ET)$_2$X in the following) is known to exhibit 
some anomalous COs~\cite{MWatanabe1,MWatanabe2} and it is suggested that the superconductivity is induced by charge fluctuation in the 
vicinity of the CO state.~\cite{HKobayashi}

$\theta$-(ET)$_2$X consists of alternating stack of X$^-$(closed shell) and ET$^{1/2+}$(3/4-filled) layers and
the latter contribute electrical conductivity. ET layer can be regarded as two-dimensional anisotropic triangular lattice with
$3/4$-filled band. In a system away from half-filling, the charge degrees of freedom become important as well as the spin degrees 
of freedom. In addition, we have to take into account the long-range Coulomb interaction because the screening effect is rather weak
in this system. To treat this problem, an extended Hubbard model which includes the nearest-neighbor Coulomb interaction $V_{ij}$ have been
studied extensively. The mean-field study by Seo~\cite{Seo} successfully reproduces the horizontal stripe CO (insulator) in X=RbZn(SCN)$_4$ 
confirmed by some experiments and supports the validity of the model. On the other hand, recent X-ray diffraction experiments have shown
the existence of non-stripe type CO in some materials. In X=RbZn(SCN)$_4$, short-range CO with wave vector 
$\boldsymbol{q}=(1/3, k, 1/4)$ is observed in a high temperature metallic region.~\cite{MWatanabe2} In X=CsCo(SCN)$_4$, the coexistence of
different COs, $\boldsymbol{q}_1=(2/3,k,1/3)$ and $\boldsymbol{q}_2=(0,k,1/2)$, is suggested.~\cite{MWatanabe1} These non-stripe type COs 
are not long-range order and only seen in a metallic region. It is considered that the large value of nearest-neighbor Coulomb interaction 
$V_{ij}$ and the geometrical frustration cause the charge fluctuation and the anomolous COs. Theoretically, several mean-field studies
\cite{TMori,Kaneko} for the extended Hubbard model suggest the existence of non-strype CO with ``3-fold" periodicity in the relevant 
parameter spaces for real materials. 

As for the superconductivity in X=I$_3$, its detailed property is still uncertain. Some theoretical studies discuss the charge fluctuation
mechanism caused by the nearest-neighbor Coulomb interaction $V$. In a square lattice, the large value of $V$ enhances the charge
fluctuation at $(\pi ,\pi)$ and it leads to the checkerboard-type CO~\cite{McKenzie,Hanasaki} and the $d_{xy}$-wave 
superconductivity.~\cite{Merino1,AKobayashi} On the other hand, in a triangular lattice, the RPA study shows that $V$ enhances the charge 
fluctuation at $(2\pi/3, 2\pi/3)$ and the triplet next-nearest-neighbor $f$-wave superconductivity is induced.~\cite{Tanaka} This result is 
quite interesting since the triplet superconductivity is hard to be realized compared with the singlet one in general. However, the above 
studies are limited to the mean-field level. Therefore, we need more accurate calculations which can treat the charge fluctuation more
correctly.          

\begin{figure}[t]
\begin{center}
\includegraphics[width=7cm]{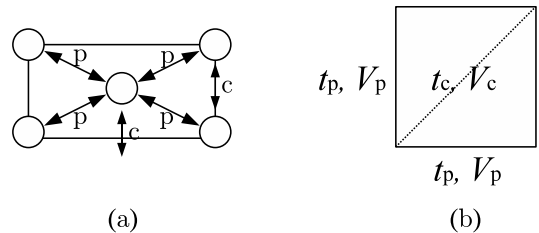}
\end{center}
\caption{(a) Schematic representation of the arrangement of donor ET molocules. p and c denote vertical and diagonal directions, 
respectively. (b) Equivalent frustrated square lattice.}
\label{f1}
\end{figure}

With these facts in mind, we study the following extended Hubbard model in a two-dimensional anisotropic triangular lattice as the 
relevant model of $\theta$-(ET)$_2$X,
\begin{equation}
 H=\sum_{\left<i,j\right>\sigma}\left(t_{ij}c^{\dagger}_{i\sigma}c_{j\sigma}+\mathrm{h.c.}\right)
  +U\sum_{i}n_{i\uparrow}n_{i\downarrow}+\sum_{\left<i,j\right>}V_{ij}n_{i}n_{j}, \label{ExtHub}
\end{equation}
where $\left<i,j\right>$ denotes the summation over the nearest-neighbor sites. As shown in Fig.~\ref{f1}(a), the values of $t_{ij}$ 
and $V_{ij}$ depend on their directions: $t_{\mathrm{p}}$ and $V_{\mathrm{p}}$ for the diagonal direction and $t_{\mathrm{c}}$ and 
$V_{\mathrm{c}}$ for the vertical direction, respectively. This lattice structure is topologically equivalent to a frustrated square
lattice as shown in Fig.~\ref{f1}(b). We set $U/t_\mathrm{p}=10$ and $t_\mathrm{c}/t_\mathrm{p}=0$ as the relevant 
values for X=CsCo(SCN)$_4$ $(|t_\mathrm{c}/t_\mathrm{p}|\ll 1)$~\cite{HMori2} and vary the values of $V_\mathrm{p}$ and $V_\mathrm{c}$. 
To investigate the ground state of this Hamiltonian, we use variational Monte Carlo (VMC) method.
We introduce the following Jastrow-type trial wave function,
\begin{equation}
 \left|\Psi \right> = P_V P_W P_G \left|\Phi \right>, 
\end{equation}
where
\begin{align}
 P_G &= g^{\sum_in_{i\uparrow}n_{i\downarrow}}, \\
 P_W &= w^{\sum_{(\mathrm{p})}n_in_j}, \\
 P_V &= v^{\sum_{(\mathrm{c})}n_in_j}.
\end{align}
$P_G$ is a Gutzwiller projection operator which reduces the probability of double occupancy of electrons at the same site. $P_W$ and $P_V$
are nearest-neighbor projection operators which control the weight of the electron configuration at the nearest-neighbor sites. 
Summations over (p) and (c) correspond to the diagonal and vertical directions, respectively. The parameters $g, w$ and $v$ are variational 
parameters with $0\leq g, w, v\leq 1$. If we set $g=w=v=1$, $\left|\Psi \right>$ is reduced to $\left|\Phi \right>$ which is constructed 
from a mean-field solution of a certain Hamiltonian such as paramagnetic metal, CO and superconductivity.

\begin{figure}[t]
\begin{center}
\includegraphics[width=7.0cm]{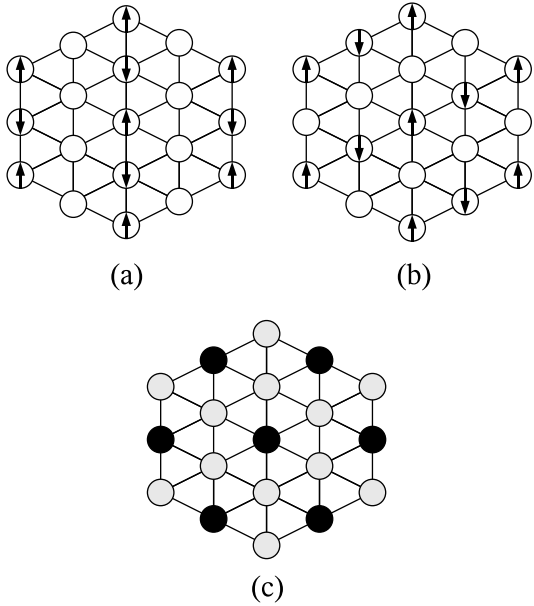}
\end{center}
\caption{Three different CO patterns studied in this paper. (a) Vertical stripe and (b) Diagonal stripe state. Circles with up or 
down arrow denote charge-rich sites and open circles denote charge-poor sites. (c) The 3-fold state where black (gray) circles represent
charge-rich A (charge-poor B and C) sublattices. }
\label{f2}
\end{figure}

For the trial wave function of CO, we consider three different patterns as shown in Fig.~\ref{f2}. We call them (a) vertical stripe, (b) 
diagonal stripe and (c) 3-fold state, respectively. When $V_\mathrm{p}\gg V_\mathrm{c}$, the vertical stripe is expected because this 
pattern fully avoids the energy loss of $V_\mathrm{p}$. For the same reason, the diagonal stripe is expected when
$V_\mathrm{p}\ll V_\mathrm{c}$. On the other hand, when $V_\mathrm{p}\simeq V_\mathrm{c}$, the possibility of non-stripe type CO exists 
since the stripe CO does not avoid both $V_\mathrm{p}$ and $V_\mathrm{c}$ simultaneously. Although we can consider other various types of 
non-stripe COs with long periodicity, we study the so-called ``3-fold'' state proposed by Mori~\cite{TMori} and discussed in a mean-field
study~\cite{Kaneko} as the first step. The 
3-fold state consists of charge-rich A sublattice and charge-poor B and C sublattices, as shown in Fig.~\ref{f2}(c). This state is always 
metallic because the charge gap opens away from the Fermi energy. It is due to the fact that the 3-fold state is incommensurate for 
the triangular lattice with $1/4$-filled band.  

For the trial wave function of superconductivity, we adopt the BCS-type wave function. The wave function with fixed electron numbers is
used for calculational convenience. The explicit form is given previously.~\cite{HWatanabe1}. 

\begin{figure}[t]
\begin{center}
\includegraphics[width=6.5cm]{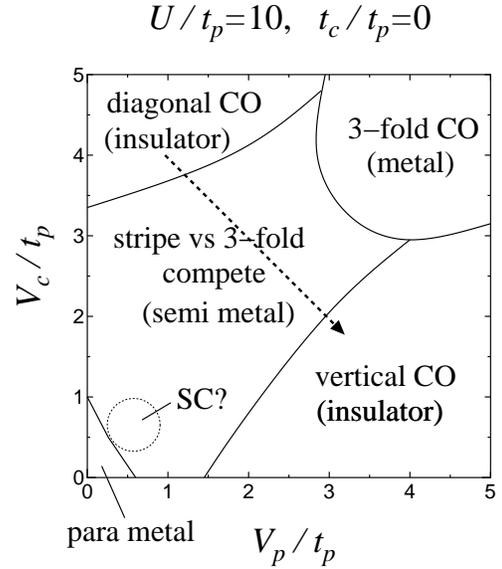}
\end{center}
\caption{The ground state phase diagram in a $V_\mathrm{p}-V_\mathrm{c}$ plane. Dashed circle represents the region where the 
superconductivity can appear. In Fig.~\ref{f4}, we show the energy comparison of three COs along the dashed arrow in the center of the 
phase diagram.}
\label{f3}
\end{figure}

\begin{figure}[t]
\begin{center}
\includegraphics[width=7.4cm]{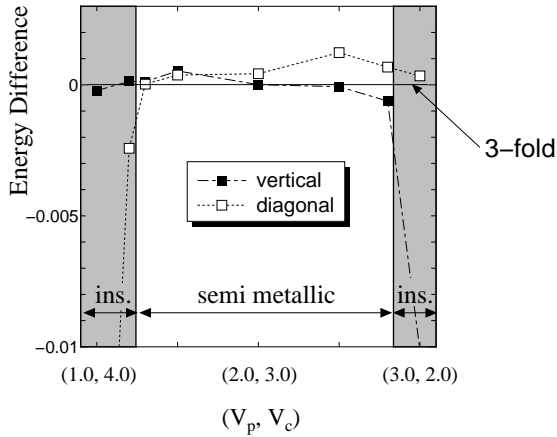}
\end{center}
\caption{The relative energy compared to the 3-fold state. The horizontal axis is along the dashed arrow shown in Fig.~\ref{f3}. 
The system is in insulating state at both end of the axis and in semi metallic state at the center.}
\label{f4}
\end{figure}

\begin{figure}[t]
\begin{center}
\includegraphics[width=7.6cm]{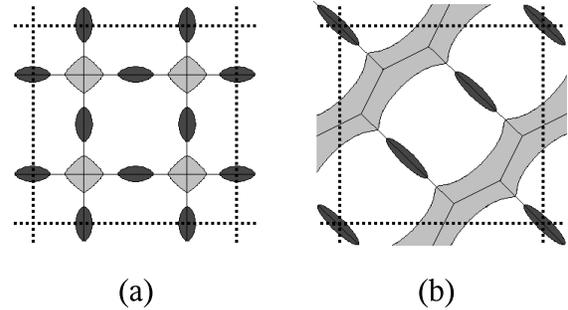}
\end{center}
\caption{Schematic representation of Fermi surfaces in the equivalent square lattice (Fig.~\ref{f1}(b)) for (a) the vertical stripe and (b)
the 3-fold state, respectively. The black (gray) regions represent electron (hole) pockets. The dotted lines represent the original first
Brillouin zone.}
\label{f5}
\end{figure}

Figure~\ref{f3} shows the phase diagram in $V_\mathrm{p}-V_\mathrm{c}$ plane obtained in the present VMC calculations. The vertical stripe 
is stable in the lower right-hand region $(V_\mathrm{p}\gg V_\mathrm{c})$ and the diagonal stripe is stable in the upper left-hand region
$(V_\mathrm{p}\ll V_\mathrm{c})$, as expected. This result is consistent with the exact diagonalization study.~\cite{Merino2}  
In the upper right-hand region, where the nearest-neighbor interactions are competing and large, the 3-fold metallic state is stable. 
This state is always metallic as mentioned before and have no spin order due to the geometrical frustration. 
In the center of the phase diagram, on the other hand, the stripes and the 3-fold state have almost comparable condensation energies as 
shown in Fig.~\ref{f4}. Interestingly, all of them are not insulating because they have semi-metallic Fermi surface with hole pockets and 
electron pockets as shown in Fig.~\ref{f5}. These ``charge-ordered metal'' states are intermediate ones caused by the frustration of 
nearest-neighbor Coulomb interaction. 

Let us here compare our results with experiments. In X=CsCo(SCN)$_4$, the coexistence of different short-range CO patterns (the horizontal 
stripe and the 3-fold type) is observed in X-ray measurements.~\cite{MWatanabe1} This means that the stripes and the 3-fold state are 
energetically competing with each other, as our result shows, and neither of them can become a long-range order at finite temperatures. 
For more accurate discussion, however, it will be necessary to consider the lattice degrees of freedom. The horizontal stripe observed in 
the experiments has higher energy than the diagonal one and does not appear in our phase diagram (Fig.~\ref{f3}). However, it is confirmed 
both experimentally~\cite{MWatanabe2} and theoretically~\cite{Seo, Kaneko} that the horizontal stripe is greatly stabilized in the 
$\theta _d$-phase caused by the lattice distortion. Indeed, the horizontal stripe insulating state is realized in X=RbZn(SCN)$_4$ 
with structural phase transtion to $\theta _d$-structure.~\cite{MWatanabe2} Thus we speculate that, in X=CsCo(SCN)$_4$, the lattice 
distortion occurs locally and the horizontal stripe is stabilized, although it cannot be a long-range order since the 3-fold state has 
alomost the same energy and prevents the growth of horizontal stripe. 

Next, we discuss the possibility of superconductivity. As shown in Fig~\ref{f3}, we find that the novel type of superconductivity has 
comparable condensation energy with those of the stripes and the 3-fold state in the lower left-hand region. Since the variational energies
are so close to each other, it is rather difficult to determine which state is most stable. The pairing symmetry is the triplet 
next-nearest-neighbor (nnn) $f$-wave. The gap function of this symmetry, $\Delta(\boldsymbol{k})$, changes its sign six times in a 
$\boldsymbol{k}$-space and it corresponds to the nnn pairing in the real space.~\cite{HWatanabe1} The reason why the nnn $f$-wave is 
enhanced is as follows. When the nearest-neighbor Coulomb interaction is introduced, electrons avoid each other and nearest-neighbor pairing
is suppressed. As a result, most stable pairing becomes possible on the nnn sites like the 3-fold state (Fig~\ref{f2}(c)). If the 
interaction is nearly isotropic, namely $V_\mathrm{p}\simeq V_\mathrm{c}$, the 3-fold type charge fluctuation at 
$\boldsymbol{q}=(2\pi/3, 2\pi/3)$ is enhanced and nnn pairing is favored. The RPA study shows the same result.~\cite{Tanaka} This 
``charge-fluctuation-induced" superconductivity is a novel one in contrast to the ``spin-fluctuation-induced" $d_{x^2-y^2}$-wave 
superconductivity in high-$T_{\mathrm{c}}$ cuprates.

In the previous paper,~\cite{HWatanabe1} we discussed the possibility of the nnn $f$-wave superconductivity in two-dimensional isotropic 
triangular lattice with electron density $n=2/3$, having the superconductivity of Na$_{0.35}$CoO$_2\cdot 1.3$H$_2$O in mind. In that case, 
the 3-fold type CO is commensurate to the triangular lattice and greatly favored. Then the nnn $f$-wave has no chance to be stabilized since 
the condensation energy of the 3-fold state is much larger. The $1/N$ expansion study by Motrunich and Lee also shows that the obtained 
$T_{\mathrm{c}}$ is too small, although the most favored pairing is the nnn $f$-wave.~\cite{Motrunich} In contrast, the stability of the 
3-fold state at $n=1/2$ (1/4-filling) studied in this paper is much weaker than the case of $n=2/3$ because of the incommensurability. We can 
say that the stable 3-fold state is ``melted'' by the electron doping. In this condition, the nnn $f$-wave has comparable condensation energy 
and becomes the candidate of the ground state. It is possible that the superconductivity observed in X=I$_3$ has nnn $f$-wave symmetry in 
the present mechanism. Although we set $t_{\mathrm{c}}/t_{\mathrm{p}}=0$ in our calculation, the value of $t_{\mathrm{c}}/t_{\mathrm{p}}$ 
in X=I$_3$ is estimated to be $\sim 1.5$.~\cite{Kondo} $t_{\mathrm{c}}/t_{\mathrm{p}}$ dependence of the stability of the nnn $f$-wave 
superconductivity is an interesting future problem.

Finally, we discuss an unified view of the family of $\theta$-(ET)$_2$X. When we replace the anion molecule X, the symmetry of the system
is kept unchanged and the change of the nearest-neighbor Coulomb interaction will be small. However, the hopping integrals change 
drastically and in particular the ratio $t_{\mathrm{c}}/t_{\mathrm{p}}$ varies approximately from $-0.5$ to $1.5$.~\cite{HMori2,Kondo}
We consider that the value of $t_{\mathrm{c}}/t_{\mathrm{p}}$ plays important roles and causes various physical properties. In the present 
paper, we study the case of $t_{\mathrm{c}}/t_{\mathrm{p}}=0$ and show the possibility of the coexistence of the stripes and the 3-fold state 
suggested in X=CsCo(SCN)$_4$. For the case of $t_{\mathrm{c}}/t_{\mathrm{p}}=-0.5$, we obtain the result (not shown here) that the stripe 
COs are greatly favored and their region in the $V_{\mathrm{p}}-V_{\mathrm{c}}$ diagram are enlarged compared with the case of 
$t_{\mathrm{c}}/t_{\mathrm{p}}=0$.~\cite{HWatanabe2} Moreover, the region where the stripes and the 3-fold state are competing disappears. 
Combining with the mean-field study which includes the effect of lattice distortion,~\cite{Kaneko} we can explain the metal-insulator 
transition with structural change observed in X=RbM'(SCN)$_4$[M'=Co,Zn] and X=TlCo(SCN)$_4$. 
For the case of $t_{\mathrm{c}}/t_{\mathrm{p}}=0.5$, on the other hand, the stripes and the 3-fold state are suppressed and the paramagnetic 
metal state becomes dominant in the $V_{\mathrm{p}}\simeq V_{\mathrm{c}}$ region.~\cite{HWatanabe2} Thus the value of 
$t_{\mathrm{c}}/t_{\mathrm{p}}$ is crucial for the stability of CO and it will correspond to the horizontal axis of the experimental phase 
diagram of $\theta$-(ET)$_2$X proposed by Mori \textit{et al}.~\cite{HMori1} 

In summary, we have studied the extended Hubbard model on the two-dimensional anisotropic triangular lattice at 1/4-filling with variational
Monte Carlo method. We have shown that for $t_{\mathrm{c}}/t_{\mathrm{p}}=0$, the stripes and the 3-fold state are energetically competing with 
each other indicating the coexistence of different CO in $\theta$-(ET)$_2$CsCo(SCN)$_4$. We also have shown that the 3-fold type 
charge fluctuation enhances the next-nearest-neighbor $f$-wave superconductivity and that it can be applicable to the superconductivity in 
$\theta$-(ET)$_2$I$_3$. The most important parameter is the value of $t_{\mathrm{c}}/t_{\mathrm{p}}$, which leads to a unified view of 
the family of $\theta$-(ET)$_2$X. 

\section*{Acknowledgment}
The authors thank Y. Tanaka, M. Kaneko, K. Kanoda and T. Mori for useful discussions. This work is supported by Grant-in-Aids from the 
Ministry of Education, Culture, Sports, Science and Technology of Japan.

\end{document}